\documentclass[%
superscriptaddress,
 amsmath,amssymb,
 aps,
pra,
]{revtex4-1}

\usepackage{graphicx}
\usepackage{dcolumn}
\usepackage{bm}
\usepackage{physics}
\usepackage{graphicx}

\begin{document}

\title{Optimal Dephasing for Ballistic Energy Transfer in Disordered Linear Chains }
\author{Yang \surname{Zhang}}
\affiliation{Department of Physics and Engineering Physics, Tulane University, New Orleans, Louisiana 70118, USA}

\author{G.~Luca \surname{Celardo}}
\affiliation{Instituto de F\'isica, Benem\'erita Universidad Aut\'onoma de Puebla, Apartado Postal J-48, Puebla 72570, Mexico}
\author{Fausto \surname{Borgonovi}}
\affiliation{Dipartimento di Matematica e
	Fisica and Interdisciplinary Laboratories for Advanced Materials Physics,
	Universit\`a Cattolica del Sacro Cuore, via Musei 41, I-25121 Brescia, Italy}
\affiliation{Istituto Nazionale di Fisica Nucleare,  Sezione di Pavia, 
	via Bassi 6, I-27100  Pavia, Italy}

\author{Lev \surname{Kaplan}}
\affiliation{Department of Physics and Engineering Physics, Tulane University, New Orleans, Louisiana 70118, USA}

\date{June 28, 2017}

\begin{abstract}
We study the interplay between dephasing, disorder, and openness on
transport efficiency in a one-dimensional chain of finite length $N$, and in particular the 
beneficial or detrimental effect of dephasing on transport. 
The excitation moves along the chain by coherent nearest-neighbor
hopping $\Omega$, under the action of static disorder $W$ and
dephasing $\gamma$. The system is open due to the coupling of the last
site with  an external acceptor system (sink), where the excitation  can be
trapped with a rate $\Gamma_{\rm trap}$, which determines the opening strength.
While it is known that dephasing can help transport in the localized
regime, here we show that dephasing can enhance energy transfer even
in the ballistic regime. 
Specifically, in the localized regime we recover previous results, 
where the optimal
dephasing is independent of the chain length and proportional to
$W$ or $W^2/\Omega$. In the ballistic regime,
 the optimal dephasing decreases as $1/N$ or
$1/\sqrt{N}$ respectively for weak and moderate static disorder. 
When focusing on the excitation starting at the beginning of
the chain,  dephasing can help excitation transfer only
above  a critical value of disorder $W^{\rm
  cr}$, which strongly depends on the opening strength $\Gamma_{\rm
  trap}$.  
 Analytic solutions are obtained for short chains. 
\end{abstract}

\maketitle
\section{Introduction}\label{sec:I} 

The optimization of excitonic and charge transport is a central problem for
building quantum devices with different functions, including sensing,
computing, and light-harvesting. 
Theoretically, the problem is challenging due to the interplay of different environments. 
Under low light intensity,  in many 
natural photosynthetic systems or in ultra-precise photon sensors, 
the single-excitation approximation is usually valid. In 
this case the system is
equivalent to a quantum network where one excitation can hop from
site to site~\cite{cao1d,superabsorb,sarovarbio,mukameldeph,mukamelspano}.  
For a realistic description of the quantum transport
problem, however, one has to consider not only the quantum coherent
evolution, but also the coupling to multiple environments. These include an external
acceptor system (sink), where the excitation can be donated and trapped, and
the coupling with a phonon bath, which can induce different types of
disorder: Static disorder (position dependent but time independent) and
noise (time-dependent disorder). 
 
From the study of natural photosynthetic
complexes~\cite{photo,photoT,photo2,photo3,schulten} has emerged
the idea that in the optimal transport regime the energy scale of the
coherent internal coupling is the same as the scale of the coupling to the
external environment. This leaves little room for perturbative
simplifications, and the analysis of the interplay of internal and
external coupling must be carried out with care. 
Another important issue is related to finite-size effects. Indeed, many
relevant natural and artificial quantum networks are made of a few
two-level systems. For instance, the FMO complex in green sulphur
bacteria, which is thought to have the role of a quantum wire, is made
of eight bacterioclorophyll a molecules~\cite{photo}.  
The
LHI and LHII~\cite{schulten1} complexes in purple bacteria are made of 32 and
16--18 molecules, respectively. 
So the infinite system size limit also cannot be used to simplify the
problem of exciton transport. 

In a recent paper by the same authors~\cite{paperone}, exciton transport in different
quantum networks was considered in the semiclassical limit, focusing
on the role of the coupling to the external acceptor system, which
can induce coherent effects such as supertransfer of the excitation
even in the presence of large dephasing. 
Here we focus our attention on the case of a linear chain of sites with
nearest-neighbor coherent hopping of the excitation. Without invoking
the semiclassical limit (where dephasing is large with respect to the coherent
nearest-neighbor coupling), we analyze here the problem of optimal
transport in the presence of dephasing, static disorder, and a
  coherent coupling to an external acceptor system. 
We focus on the role of dephasing noise (time-dependent
perturbations) in enhancing transport, defined by the average transfer time. 
It is well known that noise is not always detrimental to transport and
in some situations may enhance efficiency~\cite{lloyd,deph}.
  Specifically, noise-enhanced transport can occur whenever coherent effects, such as
  localization, subradiance, or another kind of destructive
  interference, are acting to suppress transport.

Several works in the literature aim to understand 
the parameter regime in which transport efficiency is maximized.  
Some general principles that might be used as a guide to understand
how optimal transport can be achieved have been proposed: 
Enhanced noise assisted transport~\cite{lloyd,deph}, the  Goldilocks
principle~\cite{4goldi}, and superradiance in transport~\cite{srfmo,sr2}. 

Specifically, transport in one-dimensional chains has been studied in
depth recently in the context of closed systems~\cite{cao1d,wusilbeycao,4goldi}.  
The results obtained for a one-dimensional chain of length $N$ can be summarized as
follows: In the presence of static disorder there may be a nonzero optimal
dephasing $\gamma^{\rm opt}$ for transport. If we call $W$ the strength of
static disorder and $\Omega$ the coherent nearest-neighbor hopping,
two main regimes have been identified previously. $i)$ For $W \gg \Omega$, where the localization
length $\xi \le 1$, we have
$\gamma^{\rm opt} \propto W$, independent of $\Omega$. $ii)$ For $\Omega/\sqrt{N} \ll W \ll \Omega$, where
$1 \ll \xi \ll N$, we have nonlinear dependence on the disorder strength, 
 $\gamma^{\rm opt} \propto W^2/\Omega$. Note that in both regimes the optimal dephasing is
independent of $N$. 
On the other hand, the role of the coupling to the
acceptor systems and the value of the critical disorder needed for
dephasing to help transport have not been investigated fully. 

Contrary to what one
might expect, dephasing helps transport not only in the localized
regime, when $\xi<N$, but also in the deep ballistic regime, when
$\xi \gg N$, due to a competition between the effects of static and
dynamic disorder on transport. 
This counterintuitive result can be explained considering that 
the average transfer time of the excitation to the external sink does
not depend only on how quickly the excitation spreads along the chain,
but also on how much time it spends on the last chain
site. In the ballistic regime, the probability to be on the last site
undergoes  large fluctuations in time, which can be smoothed by 
dephasing, leading to an enhanced transfer efficiency. 
Indeed, for generic initial conditions, the excitation can be
partially trapped even in the clean system, due to the structure of
the closed-system eigenstates, as discussed for example in
Ref.~\cite{wusilbeycao}. In such situations, dephasing will aid
transport even in the absence of disorder.

Throughout, we consider the scenario where the excitation traverses
the entire length of the chain, with the initial excitation placed at
one end, and the coupling to the absorber at the other. 
In this case, dephasing
can help transport only above a
critical minimal  disorder $W^{\rm cr}$ which 
depends on the coupling to the acceptor system. 
We also find that in the deep ballistic regime the optimal dephasing  $\gamma^{\rm opt}$
is not size-independent but decreases with the chain
length $N$, up to a length determined by $N \approx \xi$, where
the optimal dephasing becomes independent of $N$.

The paper is organized as follows: In Sec.~\ref{sec-model} we define the transport model, including the effects of dephasing, disorder, and coupling to the acceptor system. In Sec.~\ref{sec-23} we obtain analytic and numerical results for the simplest finite-length chains: $N=2$ and $N=3$. Then in Sec.~\ref{sec-n} we examine the behavior for general $N$ and obtain two separate delocalized regimes where the optimal dephasing displays $N$-dependent behavior. We summarize our results in Sec.~\ref{sec-conc}.

\section{Model description}
\label{sec-model}

We study the optimal dephasing for exciton energy transfer (EET) in linear chains in the presence of disorder. The time evolution of the closed system can be expressed as 
\begin{equation}\label{master}
	i\hbar\dot{\rho}(t)=\left[H_{\rm sys},\rho(t)\right]\,.
\end{equation}
The system Hamiltonian is usually expressed in the site basis as 
\begin{equation}
H_{\text{sys}}=\sum_{i=1}^{N}\hbar \omega_{i} \ket{i}\bra{i}+ \sum_{l, m}^{N} J_{lm} \ket{l}\bra{m},  
\end{equation}
where we work in the single-exciton regime, with state $\ket{i}$ representing an excitation on site $i$ only,  $\hbar\omega_{i}$ are the site energies, and $J_{lm}$ are the inter-site couplings. 
EET systems are open and connect to acceptor systems, which serve as sinks.
In this paper, we take site $N$ to be connected to the sink. The effects of the opening are conventionally addressed by augmenting the system Hamiltonian with a non-Hermitian term:
\begin{equation}
-i {\cal W}=-i\frac{\Gamma_{\text{trap}}}{2}\ket{N}\bra{N}.
\end{equation}  
This treatment of the opening and the limits of its validity have been analyzed in
Ref.~\cite{fra1}; the opening can be thought of as due to a coherent coupling of site $N$ to a continuum of states (for instance to an
infinite lead), where the continuum has a large energy band width
with respect to the energy band width of the system.

Consequently the time evolution of the reduced density matrix $\rho$ of the system will be described as 
\begin{equation}\label{master_red}
i\hbar\dot{\rho}=\left[H_{\rm sys},\rho\right]-i\left\lbrace {\cal W},\rho\right\rbrace .
\end{equation}

EET systems are subject to background noise, which results in dephasing.
We use the Haken-Strobl-Reineker (HSR) model~\cite{Haken-Strobl} to describe the dephasing behavior of the system as 
\begin{equation}
\dot{\rho}_{ij} = -\gamma (1-\delta_{ij})\rho_{ij}.
\end{equation}  
Finally, the full system dynamics can be expressed as 
\begin{equation}\label{master_f}
\dot{\rho}_{ij} = -\frac{i}{\hbar}(H_{\text{eff}}\,\rho-\rho\,H_{\text{eff}}^\dagger)_{ij}-\gamma (1-\delta_{ij})\rho_{ij},
\end{equation}
where $H_{\text{eff}}=H_{\text{sys}}-i {\cal W}$ is the effective non-Hermitian Hamiltonian of the system. 

The efficiency of EET can be measured by the total population trapped by the sink~\cite{deph,plenio2}
\begin{equation}
\eta=\Gamma_{\text{trap}} \int_{0}^{\infty} \rho_{NN}(t) \, dt  \,,
\end{equation}
and the average transfer time to the sink~\cite{lloyd}
\begin{equation}
\tau=\Gamma_{\text{trap}} \int_{0}^{\infty} t \, \rho_{NN}(t) \, dt/\eta\,.
\end{equation}
In this paper, we neglect the fluorescence effect of excitons so that $\eta=1$, and the average transfer time $\tau$ reduces to 
\begin{equation}\label{tau_int}
\tau=\Gamma_{\text{trap}} \int_{0}^{\infty} t \rho_{NN}(t) \,dt\,.
\end{equation}
We note that if decay via fluorescence is explicitly included, we have $\eta = 1/\left(1+\Gamma_{\text{fl}} \tau\right)$ for small fluorescence rate $\Gamma_{\text{fl}}$~\cite{caosilbey}, and thus minimizing the transfer time $\tau$ is equivalent to maximizing the efficiency $\eta$.

Finally, if the master equation (\ref{master_f}) is expressed in terms of the Liouville superoperator $\mathcal{L}$
\begin{equation}\label{masterL}
\dot{\rho}(t)=-\mathcal{L} \rho(t)\,,
\end{equation} 
we have
\begin{equation}
\tau=\frac{\Gamma_{\text{trap}}}{\eta}(\mathcal{L}^{-2} \rho(0))_{\text{NN}}\,.
\label{tau_liuov}
\end{equation}

In reality EET systems are often disordered. Here we consider Anderson-type disorder, with the site energies $\omega_i$  uniformly and independently distributed in the interval $\left[-W/2, W/2\right]$, where $W$ denotes the disorder strength. The disorder-averaged transfer time is then calculated as
\begin{equation}\label{integ}
\left<\tau\right>_W= \frac{1}{W^n}\int_{-W/2}^{W/2}\ldots\int_{-W/2}^{W/2} \tau(\omega_1,\omega_2,\ldots,\omega_n) \,d\omega_1\,d\omega_2 \ldots d\omega_n.
\end{equation}   

For linear chains with uniform couplings, $J_{\text{lm}}$ takes the form $ \text{J}_{\text{lm}}=\delta_{\abs{l-m},1} \Omega $,  where $\Omega$ is the coupling constant. Furthermore, in the following we choose the initial state to be $\rho(0)=\ket{1}\bra{1}$ and set $\hbar=1$ for simplicity. 

\begin{figure}
	\includegraphics[scale=.6]{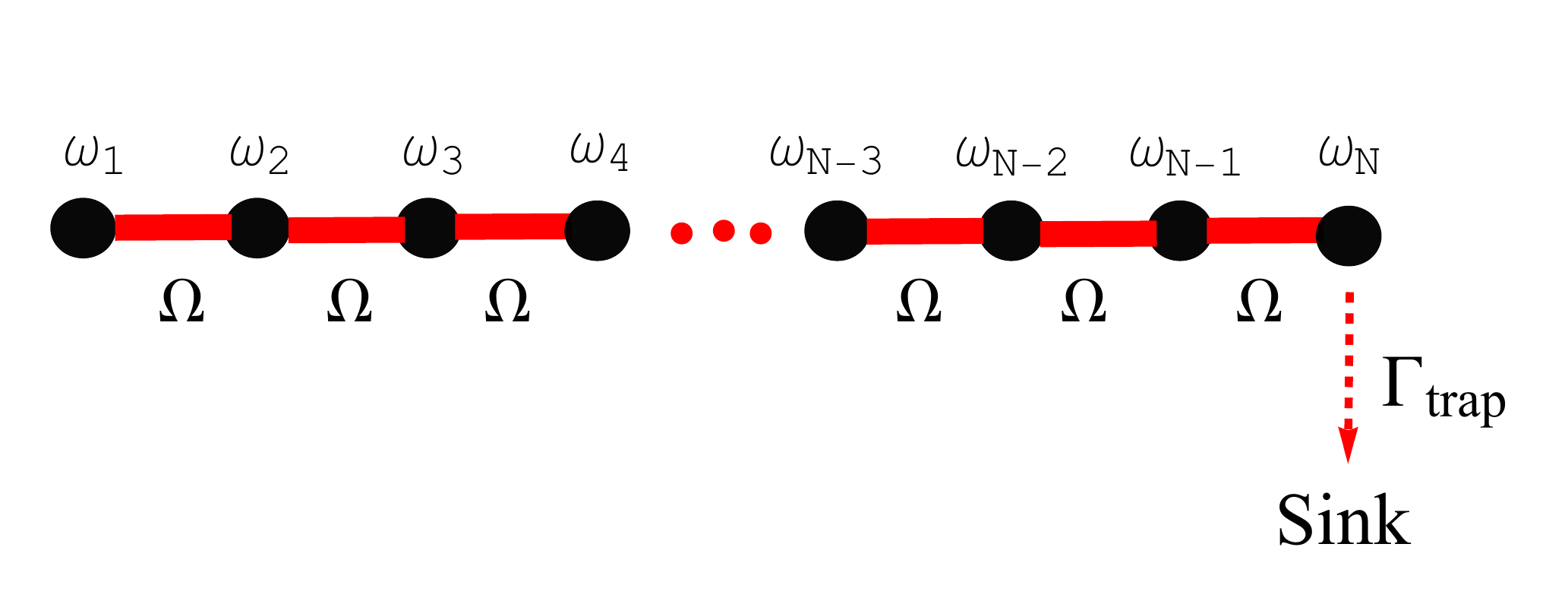}
	\caption{(Color online) A schematic of a disordered linear chain that is coupled to a sink.}
\end{figure}

\section{Optimal dephasing for two- and three-site chains}
\label{sec-23}

\subsection{Explicit Solution and Optimal Dephasing for the 2-Site Model}

For a two-site chain ($N=2$), we have obtained in Ref.~\cite{paperone} a simple analytic form for $\tau$ by solving Eq.~\eqref{tau_liuov} exactly with $\rho(0)=\ket{1}\bra{1}$:
\begin{equation}\label{tau}
\tau_2=\frac{1}{2\Omega^2}
\left( \frac{4\Omega^2}{\Gamma_{\text{trap}}}+\gamma+\frac{\Gamma_{\text{trap}}}{2}+\frac{(\omega_1-\omega_2)^2}{\gamma+ \frac{\Gamma_{\text{trap}}}{2}}\right)\,,
\end{equation}
where the subscript $2$ here and in the following denotes the chain length.
After integration over disorder using Eq.~\eqref{integ}, Eq.~\eqref{tau} becomes
\begin{equation}\label{tau2}
\left<\tau_2\right>_W=\frac{1}{2\Omega^2}
\left( \frac{4\Omega^2}{\Gamma_{\text{trap}}}+\gamma+\frac{\Gamma_{\text{trap}}}{2}+\frac{W^2}{6(\gamma+ \frac{\Gamma_{\text{trap}}}{2})}\right).
\end{equation}

We see from Eq.~\eqref{tau2} that the average transfer time behaves monotonically with static disorder strength $W$, i.e., increasing disorder always slows down transport. On the other hand, there is a complex interplay between static disorder $W$ and dephasing $\gamma$, and this interplay depends in turn on the strength of the opening $\Gamma_{\text{trap}}$. In particular, $\left<\tau_2\right>_W$ has a minimum in $\gamma$ when $W>W^{\rm cr}_2$, where 
\begin{equation} 
W^{\rm cr}_2=\sqrt{6}\Gamma_{\text{trap}}/2
\end{equation}
is the critical strength of disorder for a given degree of openness. Thus, dephasing can aid transport when disorder is sufficiently strong, $W>W^{\rm cr}_2$, and dephasing will always retard transport when $W<W^{\rm cr}_2$. In the regime $W>W^{\rm cr}_2$, the optimal rate of dephasing is given exactly by 
\begin{equation}
\gamma^{\rm opt}_2=\frac{W}{\sqrt{6}}-\frac{\Gamma_{\text{trap}}}{2}=\frac{W-W^{\rm cr}_2}{\sqrt{6}} \,.
\end{equation}

\subsection{The 3-site Chain -- Symmetry between Large and Small Opening}

\begin{figure}
	\includegraphics[scale=.6]{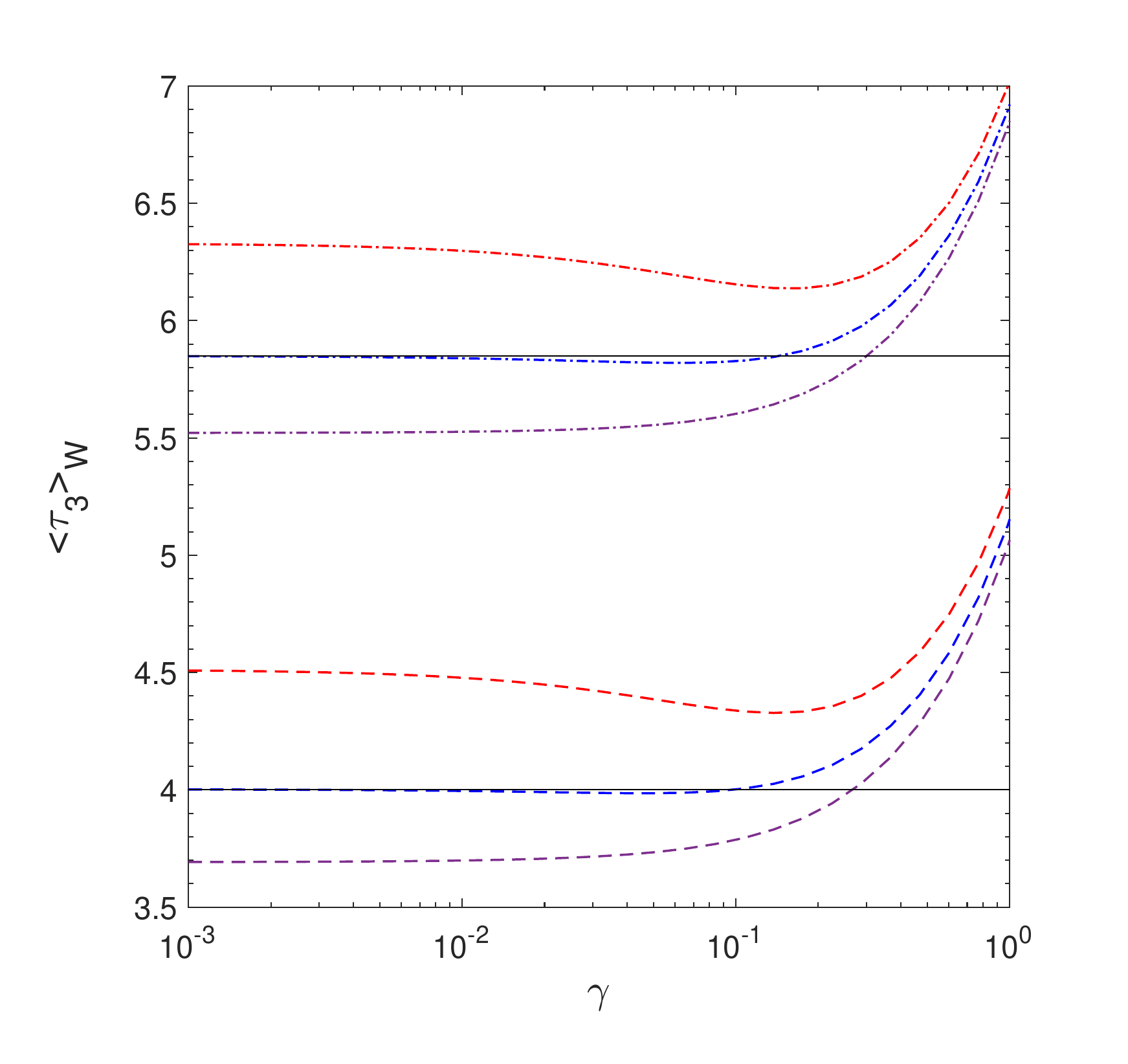}
	\caption{(Color online) Plot of the average transfer time $\left<\tau_3\right>_W$ as a function of dephasing rate $\gamma$ for a 3-site chain. Here we fix the inter-site coupling $\Omega=1$ while varying the opening $\Gamma_{\text{trap}}$ and disorder strength $W$. The three dashed curves are for opening $\Gamma_{\text{trap}}=1$, and the three dot-dashed 
	curves are for opening $\Gamma_{\text{trap}}=10$; within each group from top to bottom we have $W=1.5$, $1.1$, and $0.7$.  Physically, the presence of a minimum for $W>1$ indicates that appropriate dephasing can enhance the transport.}
	\label{fig-tau3}
\end{figure}	 

For a chain of length $N=3$, the exact transfer time $\tau_3$ may be written down explicitly for a given realization of the disorder, for any dephasing rate and any coupling strength to the sink. The result, shown in Eq.~\eqref{tau3exact} in Appendix~\ref{Appendix}, is unwieldy to work with analytically except in limiting cases; however it is easy to perform numerically the disorder integration given by Eq.~\eqref{integ}. At first glance, the behavior is qualitatively similar to that of the $N=2$ chain, as illustrated in Fig.~\ref{fig-tau3}. At fixed $\Omega=1$ and weak disorder (small $W$), the transfer time $\tau_3$ increases monotonically with dephasing rate $\gamma$, but as disorder increases,
a minimum in $\gamma$ appears and grows. The critical disorder in this case is seen numerically to be $W^{\rm cr}_3 \approx 1$ for both $\Gamma_{\text{trap}}=1$ and $\Gamma_{\text{trap}}=10$.

\begin{figure}
	\includegraphics[scale=0.7]{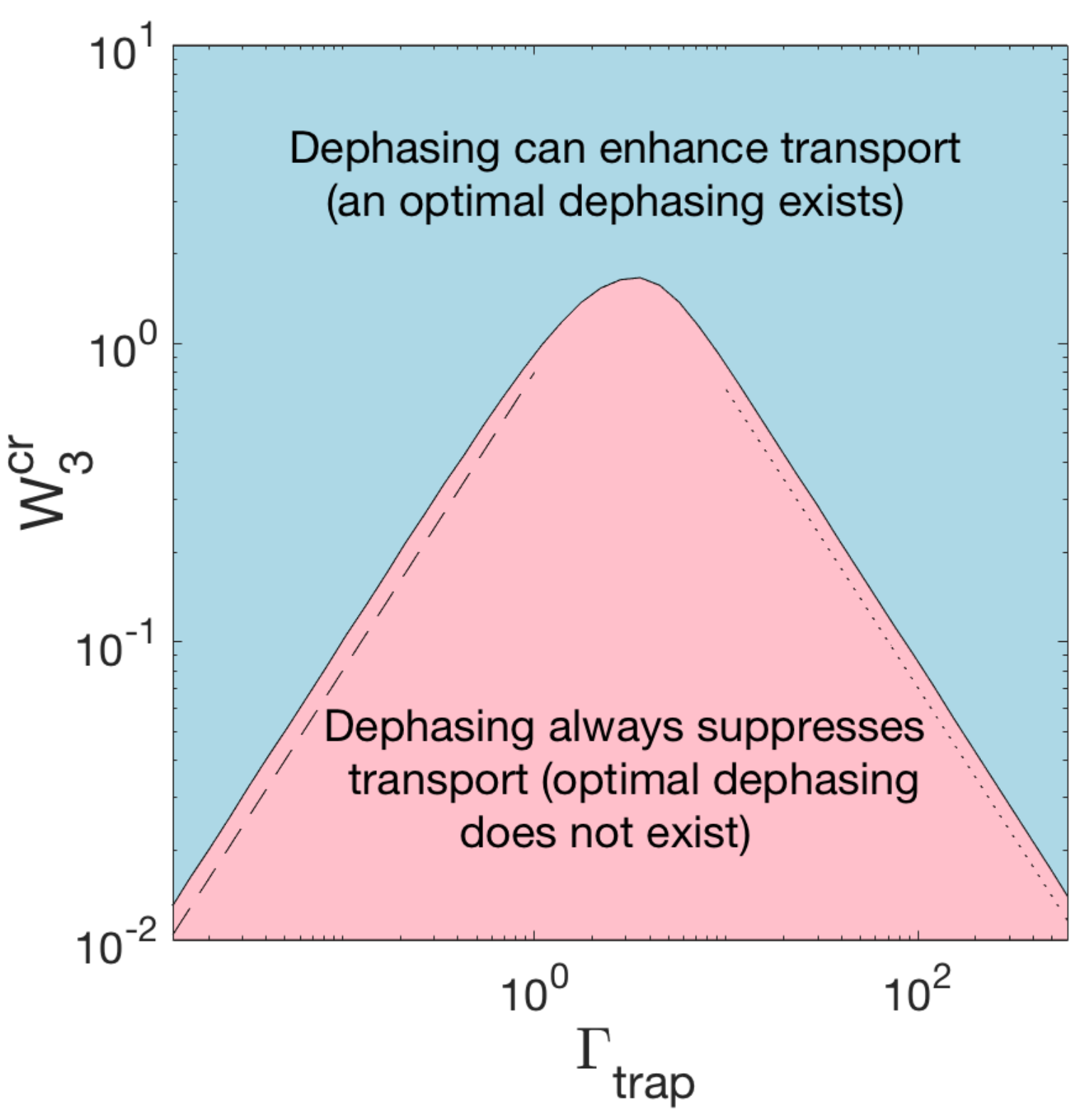}
	\caption{(Color online) Plot of critical disorder strength $W^{\rm cr}_3$ as a function of opening $ \Gamma_{\text{trap}}$ in a 3-site chain. The curve separates two transport regimes. In the upper region of the phase diagram (light blue), dephasing of the right strength will enhance transport while in the lower region (pink), dephasing always suppresses transport. The straight lines of slope +1 and -1 indicate the scaling for small and large opening $ \Gamma_{\text{trap}}$, respectively. Here we fix units where $\Omega =1 $.}\label{dia}
\end{figure}

In Fig.~\ref{dia}, for each value of the opening $\Gamma_{\text{trap}}$ we calculate the ensemble-averaged transfer time $\langle \tau_3 \rangle_W$ as a function of dephasing $\gamma$ and disorder $W$, and obtain the disorder value    
$W^{\rm cr}_3$ at which $\tau_3$ develops a minimum as a function of $\gamma$.
Looking more closely at Fig.~\ref{dia}, we find an important qualitative difference in the system's behavior as compared with the 2-site case.  For the 2-site chain, the critical disorder is always proportional to the opening strength, $W^{\rm cr}_2=\sqrt{6}\Gamma_{\text{trap}}/2$, regardless of the value of $\Omega$. Now for the 3-site chain, the critical disorder is again proportional to $\Gamma_{\text{trap}}$ for small opening, but for large opening the critical disorder {\it decreases} with the opening strength. (More precisely, we have $W^{\rm cr}_3=\Gamma_{\text{trap}} $ for $\Gamma_{\text{trap}} \ll \Omega $ and $W^{\rm cr}_3=6\sqrt{2}\Omega^2/\Gamma_{\text{trap}} $ for $\Gamma_{\text{trap}} \gg \Omega $, as will be obtained analytically below.)

The symmetry between large and small opening is related to the superradiance transition in open quantum systems, where for sufficiently large opening a segregation of resonances into superradiant states (strongly coupled to the sink) and subradiant states (trapped away from the sink) occurs, with the result that escape to the sink is suppressed~\cite{Zannals,kaplan}. Transport to the sink in the clean quantum system  is maximized at the superradiance transition.
In the case of the clean $N-$site chain (with no disorder or dephasing), the transfer time 
is given by
\begin{equation}
\tau=\frac{N}{\Gamma_{\text{trap}}}+\frac{(N-1)\Gamma_{\text{trap}}}{4\Omega^2}\,,
\label{tauclean}
\end{equation}
see, e.g., Ref.~\cite{paperone},
and thus transport to the sink is optimized  at $  \Gamma_{\text{trap}} = 2 \sqrt{N/(N-1)}\Omega $, or  $\Gamma_{\text{trap}} = \sqrt{6} \Omega $ for $N=3$.

\subsubsection{Analytics for Small Opening}
\label{small3}

We now obtain analytically the critical disorder for the 3-site chain. We break up the problem into two regimes, starting with the regime of weak opening: $\Gamma_{\text{trap}} \ll \Omega$.
After expanding $\tau_3$ from Appendix \ref{Appendix} in powers of $\Omega^{-1}$ assuming $\Omega$ is large compared to all other energy scales in the problem, and integrating the expanded expression over disorder using Eq.~\eqref{integ}, we obtain
\begin{equation}\label{tau3}
\left<\tau_3\right>_W= \frac{3}{\Gamma _{\text{trap}}}+\Omega^{-2} \left( \frac{3 \gamma }{2}+\frac{\Gamma _{\text{trap}}}{2}-\frac{W^2}{12 \left(2 \gamma +\Gamma
		_{\text{trap}}\right)}+\frac{5 W^2}{12 \left(4 \gamma +\Gamma _{\text{trap}}\right)}\right)+O\left(\Omega^{-4}\right).
\end{equation} 
Differentiating $\left<\tau_3\right>_W$ with respect to $\gamma$, and neglecting $O(\Omega^{-4})$ terms, we find
\begin{equation}\label{dtdg}
\frac{\partial \left<\tau_3\right>_W }{\partial\gamma} \approx \frac{1}{12 \Omega ^2} \left[18+ W^2 \left(\frac{2}{\left(2 \gamma +\Gamma _{\text{trap}}\right){}^2}-\frac{20}{\left(4
	\gamma +\Gamma _{\text{trap}}\right)^2}\right)\right] \,.
\end{equation}
Now $\frac{2}{\left(2 \gamma +\Gamma _{\text{trap}}\right){}^2}-\frac{20}{\left(4
	\gamma +\Gamma_{\text{trap}}\right)^2} $ is always negative for non-negative $\gamma$ and $\Gamma_{\text{trap}}$.
Furthermore, the quantity in square brackets in Eq.~(\ref{dtdg}) increases
monotonically from $18(1-W^2/\Gamma_{\text{trap}}^2)$ to 18 as $\gamma$ increases from 0 to $\infty$. Thus for $W<\Gamma_{\text{trap}} $, Eq.~\eqref{dtdg} is always positive, and dephasing always retards transport. For $W>\Gamma_{\text{trap}}$, on the other hand, Eq.~\eqref{dtdg} increases monotonically in $\gamma$ from below 0 to above 0, i.e., the mean transfer time $\left<\tau_3\right>_W$ exhibits a minimum as a function of $\gamma$. Thus, the critical disorder strength for weak opening is given by 
\begin{equation}
W^{\rm cr}_3=\Gamma_{\text{trap}} \,.
\end{equation}

What about the optimal dephasing $\gamma^{\rm opt}_3$? As a function of disorder strength $W$ when $W>W^{\rm cr}_3$, this is given in general by the solution of a quartic equation. Nevertheless, three relatively simple regimes may be distinguished, the first two of which fall within the range of validity of the large-$\Omega$ approximation, Eq.~\eqref{tau3}.

(i) For weak disorder only slightly above the critical value, $ 0<W-W^{\rm cr}\ll W^{\rm cr}$, we may expand Eq.~(\ref{dtdg}) and obtain $\gamma^{\rm opt}_3\approx \frac{9}{38}(W-W^{\rm cr}_3)$. 

(ii) For moderate disorder,  
$W^{\rm cr} \ll W \ll \Omega$, $\gamma^{\rm opt}$ will be large compared to $W^{\rm cr}_3=\Gamma_{\text{trap}}$, and thus we may take $\gamma \gg \Gamma_{\text{trap}}$ in Eq.~(\ref{dtdg}). We then have $\gamma^{\rm opt}_3\approx W/2\sqrt{6} $.

 (iii) Finally, one may consider the behavior for strong disorder, $W^{\rm cr}_3 \ll \Omega \ll W$. This is outside the range of validity of the above derivation, since $\Omega$ is no longer the largest energy scale. In this parameter regime, to be discussed further in Sec.~\ref{optgammaregimes}, the optimal dephasing rate converges to the $N$-independent form $\gamma^{\rm opt} \approx W/\sqrt{6}$.
 
 Interestingly, in each of the three ranges of the disorder strength $W$, the optimal dephasing rate $\gamma^{\rm opt}_3$ grows linearly with $W$, but the coefficient is different in each case. In Sec.~\ref{optgammaregimes}, we will see that each of the three regimes identified here for $N=3$ has a counterpart at large $N$, but each is associated with a different scaling with system size $N$.

\subsubsection{Analytics for Large Opening}
\label{large3}

Now we consider the strong opening scenario, $\Gamma_{\text{trap}}\gg \Omega $. As far as the scaling analysis is concerned, a large opening  can be thought as a small opening with the effective small opening strength $\Gamma_{\text{trap}}'=\Omega^2/\Gamma_{\text{trap}} $. So starting from the exact expression for $\tau_3$ in Appendix \ref{Appendix}, we may change variables from $\Gamma_{\text{trap}}$ to $\Gamma_{\text{trap}}'$, and expand $\tau_3$ assuming $\Omega$ is very large (compared to all $\omega_i$, $\gamma$, and $\Gamma_{\text{trap}}'$). Integrating over disorder, we obtain
 \begin{equation}
\left<\tau_3\right>_W'=\frac{1}{2 \Gamma _{\text{trap}}' }+\frac{18 \left(\gamma +2 \Gamma
	_{\text{trap}}'\right)+\frac{W^2}{\gamma +2 \Gamma _{\text{trap}}'}}{\Omega ^2}+O\left( \frac{1}{\Omega ^4}\right).
 \end{equation}
Straightforward algebra now shows that for strong opening  the critical disorder is given by 
\begin{equation}
W^{\rm cr}_3=6\sqrt{2}\Gamma _{\text{trap}}'=6\sqrt{2}\Omega^2/\Gamma_{\text{trap}}\,,
\end{equation}
 and the optimal dephasing rate above critical disorder is seen to be 
 \begin{equation}
 \gamma^{\rm opt}_3=(W-W^{\rm cr}_3)/3\sqrt{2}\,.
 \end{equation}
 
 \begin{figure}
 	\includegraphics[scale=.8]{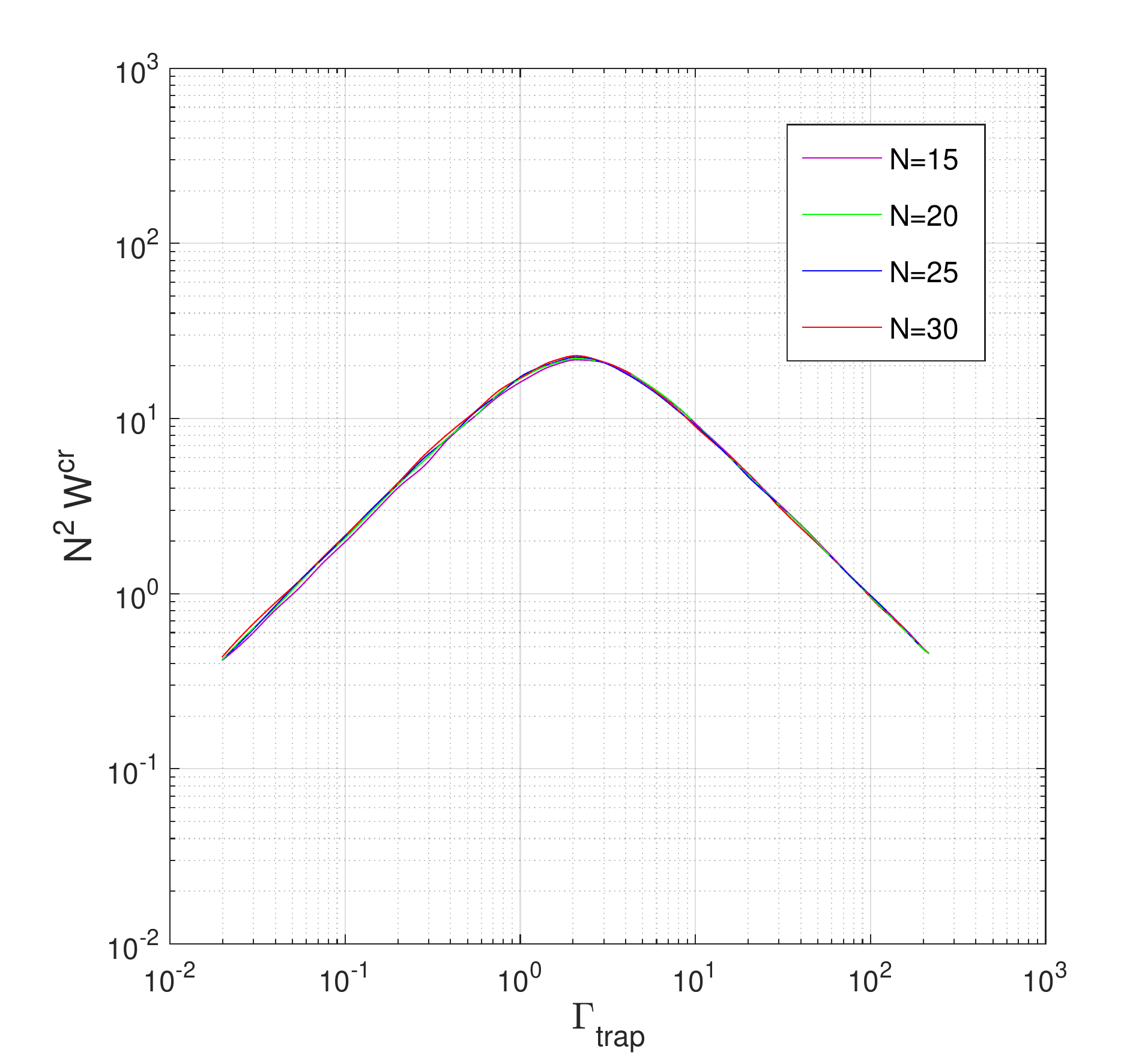}
 	\caption{(Color online) Plot of the rescaled critical disorder $ \text{N}^2 \text{W}^{\rm cr}$ as a function of opening $ \Gamma_{\text{trap}}$ for various chain lengths $N$.  Here we fix $\Omega =1$.}
 	\label{diagram_n}
 \end{figure}
 
\section{Optimal Dephasing for Long Chains}
\label{sec-n}
\subsection{Critical Disorder Strength for Long Chains}

We now consider how the results obtained above for 2- and 3-site chains may extend to chains of general length $N$. To begin with, we generalize the results of Fig.~\ref{dia} to $N$ sites. Once again, without loss of generality we choose units where hopping $\Omega=1$ and evaluate numerically, as a function of opening $\Gamma_{\text{trap}}$, the critical disorder $W^{\rm cr}$ at which the ensemble-averaged transfer time $\langle \tau(\gamma) \rangle_W$ develops a minimum as a function of dephasing rate $\gamma$. For general $N$, Monte Carlo integration is used to evaluate the disorder average. The results, for selected values of $N$, are shown in Fig.~\ref{diagram_n}.  We notice in Fig.~\ref{diagram_n} the same qualitative behavior observed earlier in Fig.~\ref{dia} for the 3-site chain. Furthermore, we see empirically that the behavior becomes $N$-independent for large $N$ when the rescaled disorder strength $N^2 W^{\rm cr}$ is plotted as a function of opening $\Gamma_{\text{trap}}$, indicating that the critical
disorder scales as
\begin{equation}
W^{\rm cr} \sim \frac{1}{N^2}
\label{WscalingN}
\end{equation} 
for all values of $\Gamma_{\text{trap}}$. In particular, comparing with the results for $N=3$, we have
\begin{equation}
W^{\rm cr} \sim \frac{\Gamma_{\text{trap}}}{N^2}
\end{equation}
for small opening, $\Gamma_{\text{trap}} \ll \Omega$, and
\begin{equation}
W^{\rm cr} \sim \frac{\Omega^2}{N^2\Gamma_{\text{trap}}}
\end{equation}
for large opening, $\Gamma_{\text{trap}} \gg \Omega$.

We notice that $ W^{\rm cr} $ approaches zero as the chain length $N$ goes to infinity. This is consistent with the fact that for an infinitely long chain, the system is localized at arbitrarily weak disorder, and any amount of dephasing can break the localization, thus aiding transport.  

Unfortunately, an analytic understanding of the empirical scaling behavior (\ref{WscalingN}) is not presently available; the analysis would require a non-perturbative treatment of the effect of the opening $\Gamma_{\text{trap}}$, since near critical disorder $\Gamma_{\text{trap}}$ will be comparable to both disorder $W$ and dephasing strength $\gamma$.

\subsection{Optimal dephasing as a function of disorder}
\label{optgammaregimes}

\begin{figure}
	\includegraphics[scale=.62]{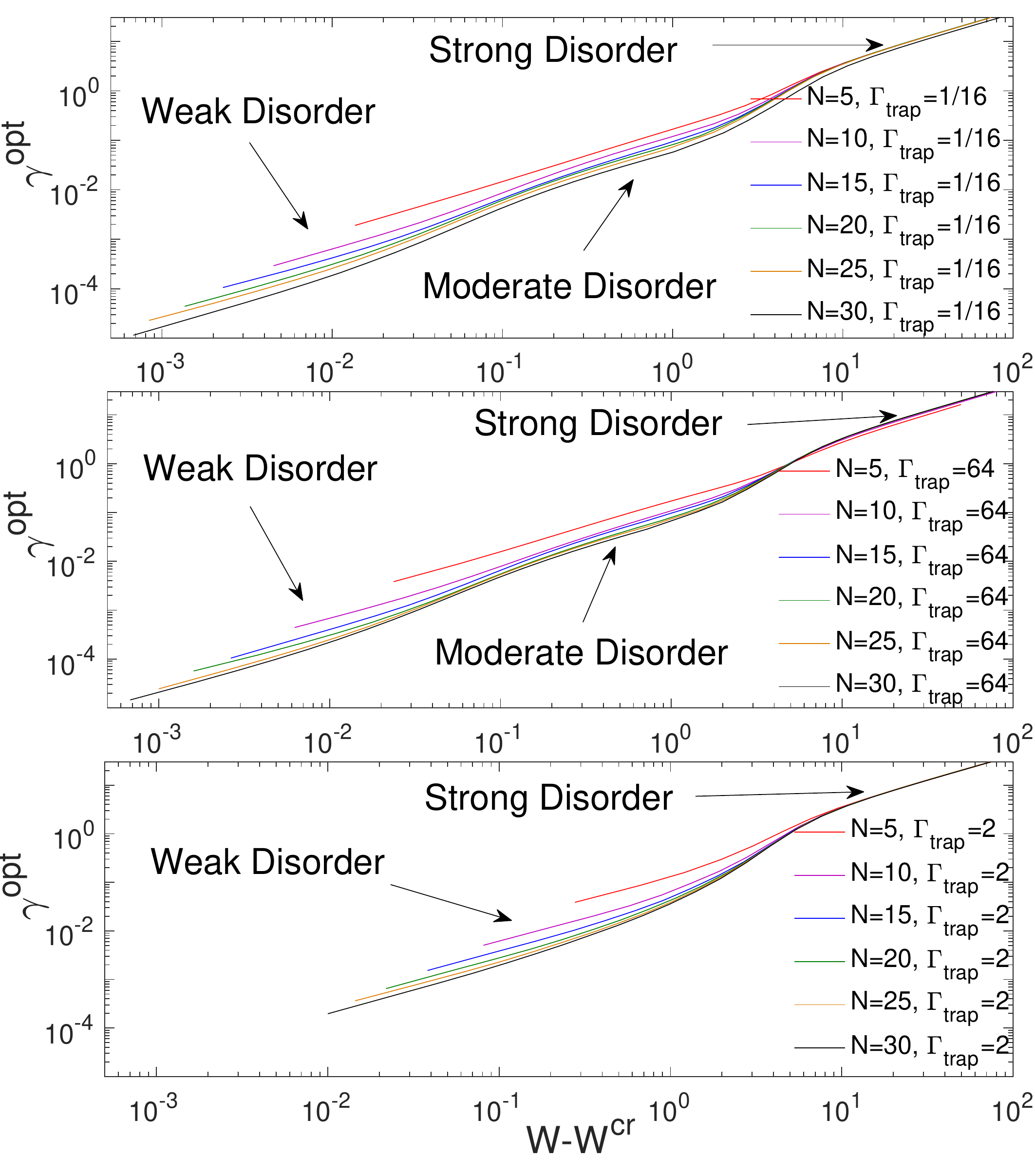}
	\caption{(Color online) Optimal dephasing rate $\gamma^{\rm opt}$ is plotted as a function of $W-W^{\rm cr}$ for chains of different length $N$, where in each curve the minimum value of $W$ is $1.4 W^{\rm cr}$ and the critical disorder $W^{\rm cr}$ is itself a function of $N$. Here we again fix $\Omega=1$. The top and middle panels show results for $\Gamma_{\text{trap}}=1/16$ and $64$, providing examples respectively of the small-opening and large-opening wings in Fig.~\ref{diagram_n}. In each case, three distinct regimes may be observed for weak, moderate, and strong disorder, which display different scaling behavior with $N$ and are analyzed in Secs.~\ref{secweak}, \ref{secmod}, and \ref{secstrong}, respectively.
		The lower panel shows results at the superradiance transition, $\Gamma_{\text{trap}}=2$; here the moderate disorder regime is absent.}  
	\label{gamma_opt}
\end{figure}

We now consider the optimal dephasing for long chains when $W>W^{\rm cr}$. Numerical results for three values of $\Gamma_{\text{trap}}$ (one corresponding to a small opening, another to a large opening, and the third to an opening right at the superradiance transition), are shown in Fig.~\ref{gamma_opt}. We see that the optimal dephasing increases monotonically with the disorder strength. However, several distinct parameter regimes can be identified, which are in direct correspondence with the three regimes obtained for $N=3$ in Sec.~\ref{small3}. Notably, each regime shows its own scaling behavior with the chain length $N$, even though this scaling is not clearly visible in Fig.~\ref{gamma_opt}. We now proceed with an analysis of the different regimes and their scaling behaviors.

\subsubsection{Weak Disorder: $W-W^{\rm cr} \sim  W^{\rm cr}$}
\label{secweak}
\begin{figure}
	\includegraphics[scale=.9]{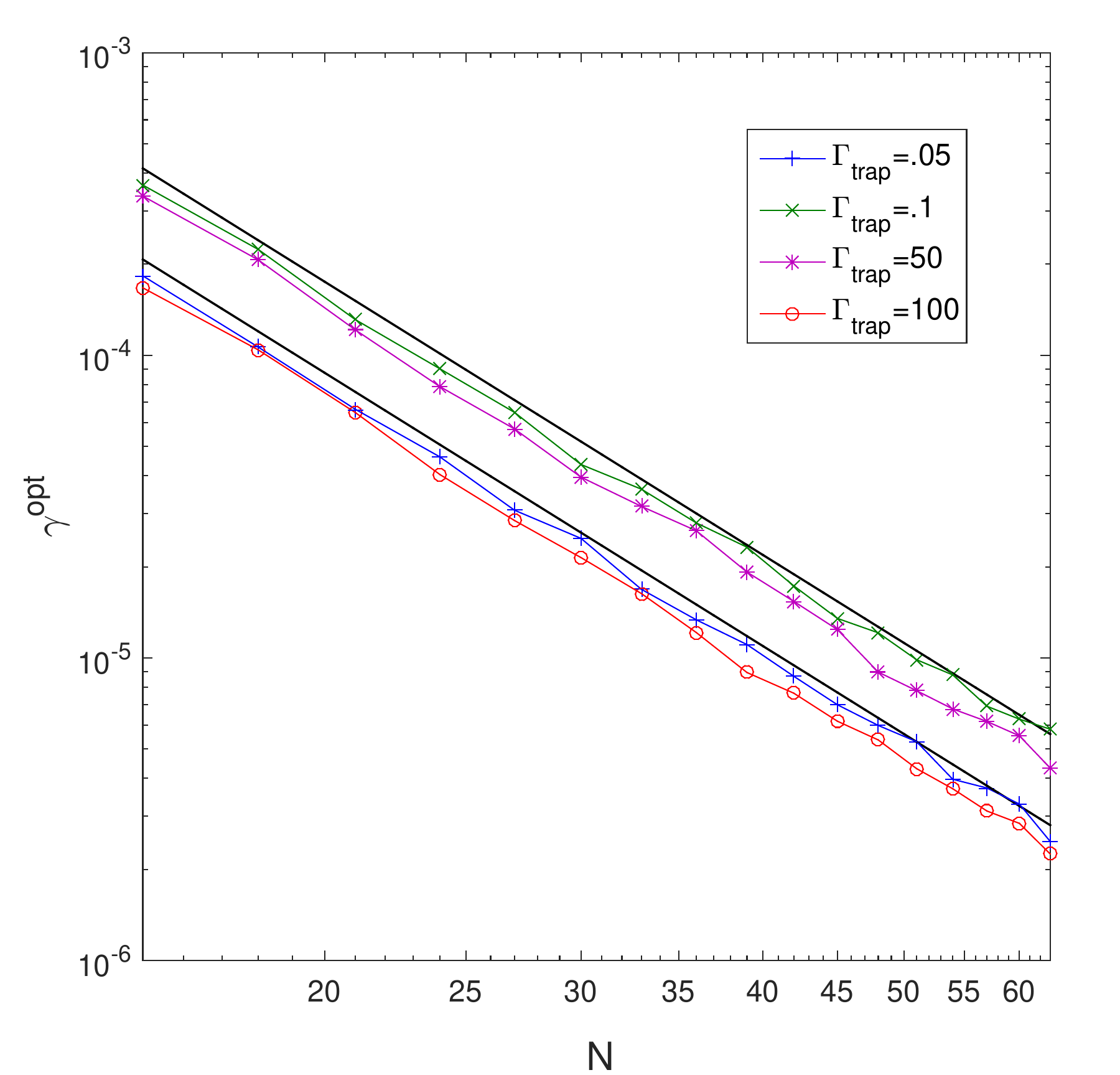}
	\caption{(Color online) Optimal dephasing rate $\gamma^{\rm opt}$ is shown as a function of N with $W=2W^{\rm cr}$ for several values of the opening strength $\Gamma_{\text{trap}}$. Here $\Omega=1$. The two black solid lines illustrate scaling proportional to $1/N^3$, implying $\gamma^{\rm opt}\sim (W-W^{\rm cr})/N$ for $W$ close to $W^{\rm cr}$.}\label{left_regime}
\end{figure}

We first consider $W$ just slightly above the critical disorder, $0<W-W^{\rm cr} \sim  W^{\rm cr}$. As seen in Fig.~\ref{gamma_opt}, here the optimal disorder $\gamma^{\rm opt}$ grows linearly with $W-W^{\rm cr}$, just as it does for $N=2$ and $N=3$. To ascertain the $N$-dependence for long chains, in Fig.~\ref{left_regime} we study $\gamma^{\rm opt}$ as a function of $N$ for $W=2W^{\rm cr}$ and several (large and small) values of the opening strength $\Gamma_{\text{trap}}$. We observe the scaling $\gamma^{\rm opt} \sim 1/N^3$ when other parameters are held fixed, which combined with Eq.~\eqref{WscalingN} implies \begin{equation}
\gamma^{\rm opt} \sim \frac{W-W^{\rm cr}}{N}\,.
\label{weakW}
\end{equation}

\subsubsection{Moderate Disorder: $\Gamma_{\text{\rm trap}}/\sqrt{N} \ll W \ll \Omega/\sqrt{N}$}
\label{secmod}
Here we  consider the behavior where disorder (and dephasing) are strong compared to the opening size but still weak compared to the hopping amplitude. Thus, we are interested in the regime $\Gamma_{\text{trap}} \ll W \sim \gamma \ll \Omega$ where any required $N$ dependence is temporarily omitted from the inequalities.

\begin{figure}
	\includegraphics[scale=0.9]{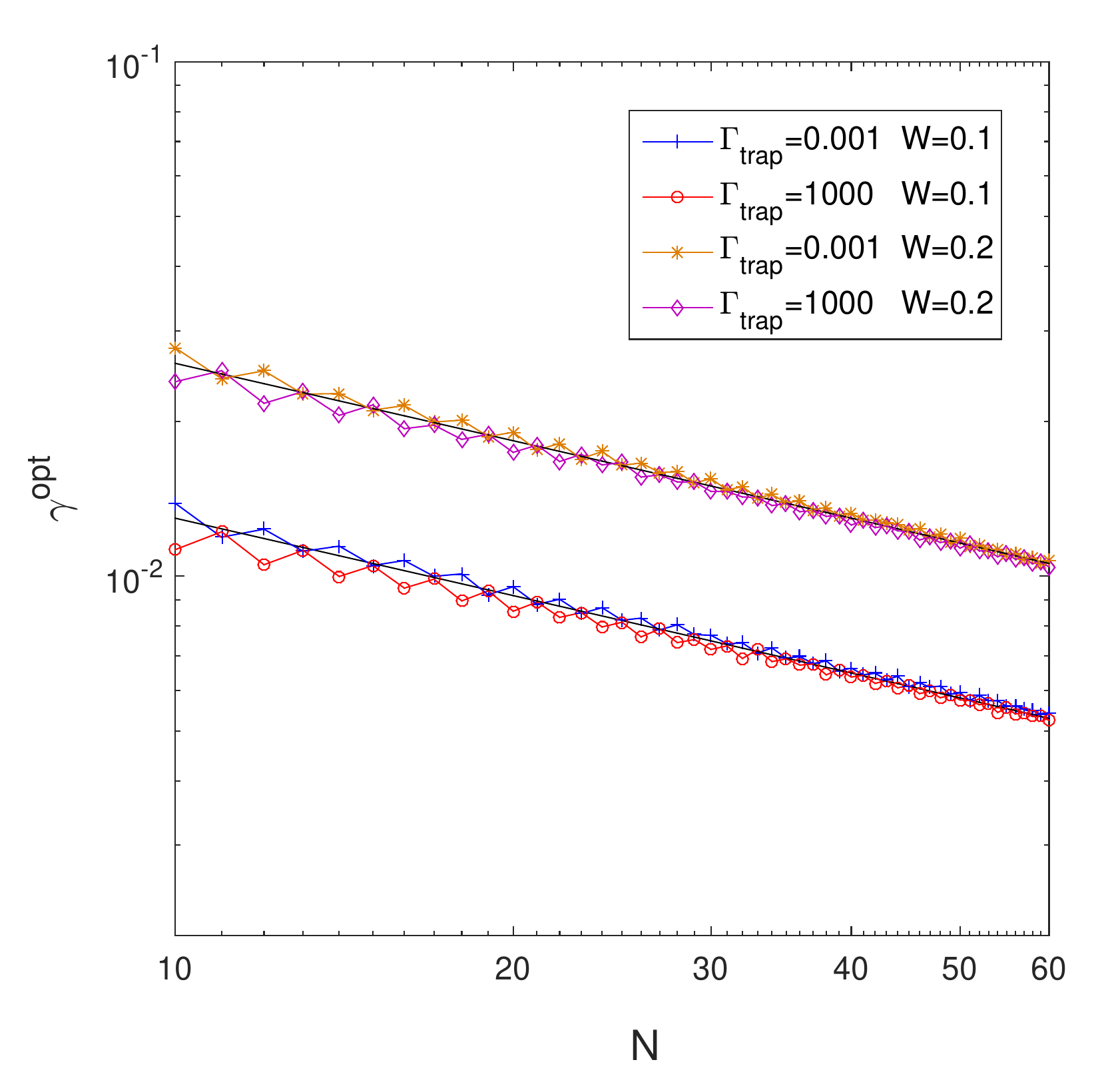}
	\caption{(Color online) Optimal dephasing $\gamma^{\rm opt}$ is shown as a function of chain length $N$ in the moderate-disorder regime, for several values of opening $\Gamma_{\text{trap}}$ and disorder strength $W$. Here $\Omega=1$. We observe good agreement with Eq.~\eqref{midreg}, as shown by the solid lines.}\label{middle}
\end{figure}

It is convenient to begin with a clean open chain in the presence of dephasing. Here the transfer time may be obtained
exactly as
\begin{equation}
\tau=\frac{N}{\Gamma_{\text{trap}}}+\frac{N(N-1)\gamma+(N-1)\Gamma_{\text{trap}}}{4\Omega^2}\,,
\label{taucleangam}
\end{equation}
to be compared with Eq.~(\ref{tauclean}) for the special case $\gamma=0$. Now we consider expanding in both disorder strength $W$
and opening $\Gamma_{\text{trap}}$ assuming that the opening is small compared to $\Omega$ (i.e., we work in a regime analogous to that considered in Sec.~\ref{small3} for $N=3$; an analogous treatment for $\Gamma_{\text{trap}} \gg \Omega$ may be considered as in Sec.~\ref{large3}). Beginning with Eq.~(\ref{taucleangam}) for $W=0$ and comparing with the results (\ref{tau2}) and (\ref{tau3}) for $N=2$ and $3$ respectively, we conjecture that for large $N$ the expansion takes the form
\begin{eqnarray}\label{largeNtau}
\left<\tau\right>_W &=&
\frac{N}{\Gamma
	_{\text{trap}}}+\frac{1}{\Omega^2}\left[\left(\frac{N(N-1)}{4}\gamma +a(N) \frac{W^2}{\gamma }\right)
	+\Gamma_{\text{trap}}\left( \frac{N-1}{4}-b(N) \frac{W^2}{\gamma^2}\right)+O\left(\Gamma
	_{\text{trap}}^2\right) \right] \nonumber \\
	&+& \frac{1}{\Omega^4}\left[c(N)W^2 \gamma+d(N)\frac{W^4}{\gamma} +O(\Gamma_{\text{trap}})\right]
+O\left(\frac{1}{\Omega^6}\right)\,.
\end{eqnarray}
Numerically, we find $a(N)=a_0 N$, $b(N)=b_0$, $c(N)=c_0 N^3$, and $d(N)=d_0N^2$ for large $N$. In particular, $a_0 \approx 0.042$.

For sufficiently large $\Omega$ and small $\Gamma_{\text{trap}}$ we may restrict our attention to the term $\left(\frac{N(N-1)}{4}\gamma +a_0 N\frac{W^2}{\gamma }\right)/\Omega^2$ only, which implies that
 the optimal dephasing in this regime should behave as
\begin{equation}\gamma^{\rm opt} \approx \frac{2\sqrt{a_0}\,W}{\sqrt{N}} \approx 0.41 \frac{W}{\sqrt{N}} \,.
\label{midreg}
\end{equation}
This predicted behavior with system size $N$ for moderate disorder strength is confirmed in Fig.~\ref{middle}. We observe in Fig.~\ref{middle} that while Eq.~\eqref{midreg} was obtained in the context of $\Gamma_{\text{trap}} \ll \Omega$, the same scaling behavior, $\gamma^{\rm opt} \sim 1/\sqrt{N}$, holds also for $\Gamma_{\text{trap}} \gg \Omega$ where the effective opening $\Omega^2/\Gamma_{\text{trap}}$ is small. 
 
Now to understand the range of validity of Eq.~\eqref{midreg}, we need to take a closer look at the two expansions in Eq.~\eqref{largeNtau}. On the one hand, our approximation breaks down for small disorder and dephasing when the terms proportional to $\Gamma_{\text{trap}}$
become comparable to the $\Gamma_{\text{trap}}$-independent terms we have been considering. This occurs when $\gamma^{\rm opt} \sim \Gamma_{\text{trap}}/N$, or equivalently $W \sim \Gamma_{\text{trap}}/\sqrt{N}$. On the other hand, the approximation also breaks down for larger disorder (and dephasing), when the $1/\Omega^4$ contribution becomes comparable to that of the $1/\Omega^2$ terms in the expansion. This occurs when $W \sim \Omega/\sqrt{N}$, which not coincidentally corresponds to the localization border where the localization length near the middle of the energy band, $\xi \approx 100 \,\Omega^2/W^2$~\cite{tessieri}, becomes comparable to the chain length $N$.

Thus the moderate-disorder regime in which the scaling of the optimal dephasing rate is given by Eq.~(\ref{midreg}) extends over the range $\Gamma_{\text{trap}}/\sqrt{N} \ll W \ll \Omega/\sqrt{N}$. We note that in the moderate-disorder regime as well as in the weak-disorder regime, the eigenstates are delocalized and wave packet motion is ballistic. Nevertheless, in both regimes we have shown that dephasing will aid transport.

\begin{figure}
	\includegraphics[scale=0.7]{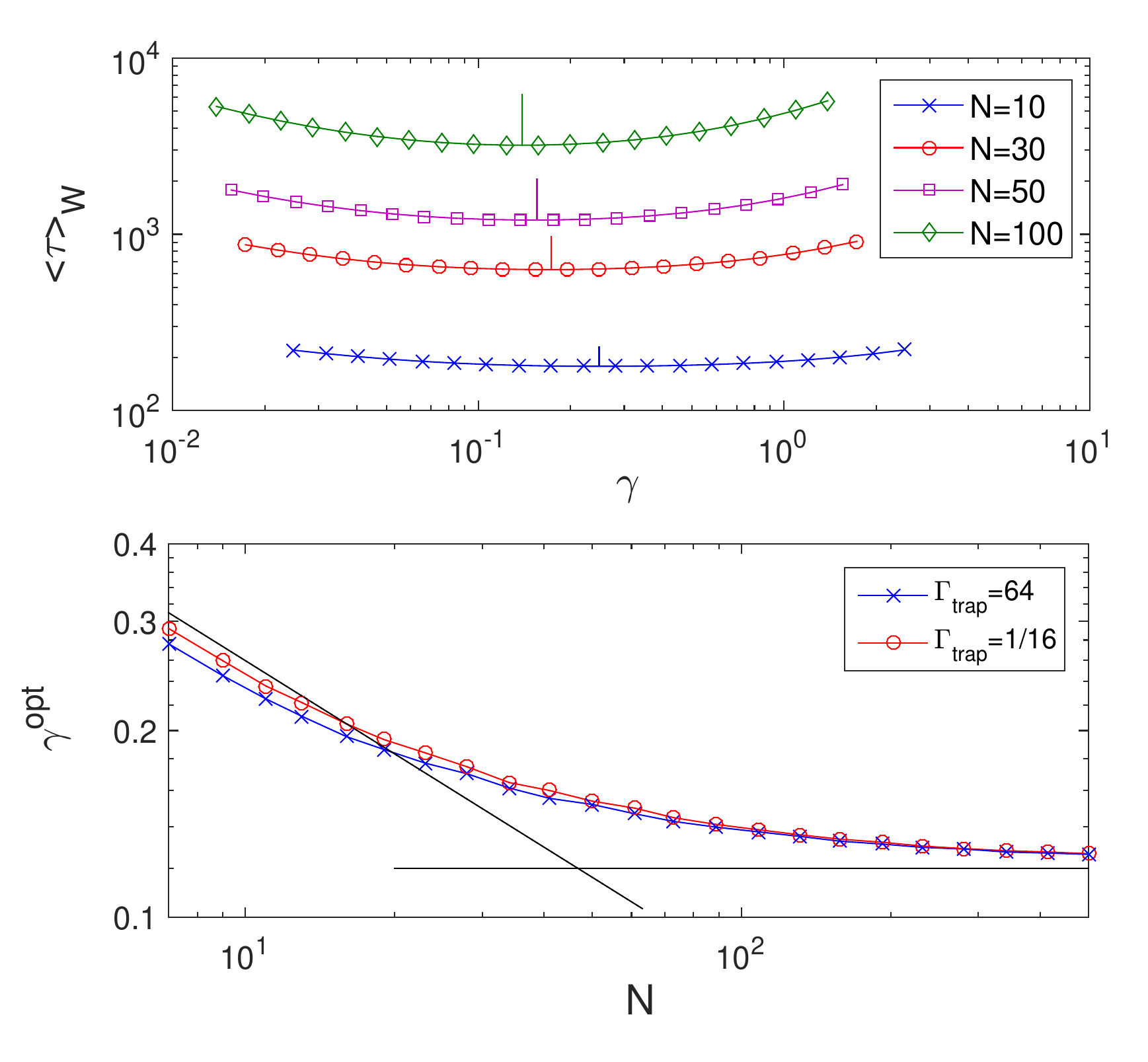}
	\caption{(Color online) Upper panel: The disorder-averaged transfer time $\langle \tau \rangle_W$ is shown as a function of dephasing rate $\gamma$ for chains of several lengths $N$. Here $\Omega=1$, $W=2$, and $\Gamma_{\text{trap}}=1/16$. In each case, the vertical line indicates the optimal dephasing rate $\gamma^{\rm opt}$. Lower panel:
The optimal dephasing $\gamma^{\rm opt}$ is shown as a function of chain length $N$ in the crossover between the moderate-disorder and strong-disorder regimes. Here $\Omega=1$, $W=2$, and two values of the opening $\Gamma_{\text{trap}}$ are
presented, corresponding to the weak-opening and strong-opening scenarios. The solid lines represent
Eq.~\eqref{midreg} in the moderate-disorder regime (which displays $N^{-1/2}$ scaling) and the $N$-independent behavior expected for strong disorder.} \label{largeN}
\end{figure}

\subsubsection{Strong disorder:  $W \gg \Omega/\sqrt{N}$ }
\label{secstrong}
Finally, in the strong disorder regime, defined by $W \gg \Omega/\sqrt{N}$, the quantum eigenstates are localized, and the dynamics is diffusive. This regime has previously been studied in the context of closed systems in Refs.~\cite{cao1d,wusilbeycao,4goldi}. More precisely, this regime comprises two sub-regimes: For $\Omega/\sqrt{N} \ll W \ll \Omega$, one has $1 \ll \xi \ll N$, and the optimal dephasing rate is given by $\gamma^{\rm opt} \sim \Omega/\xi \sim W^2/\Omega$. Upon further increase of the disorder, we reach $W \gg \Omega $, implying a localization length $\xi \sim 1$, and the optimal dephasing is then simply proportional to the disorder:
$\gamma^{\rm opt} \sim W$. Throughout the strong-disorder regime, the optimal dephasing is controlled by motion on the scale of a localization length, and as a consequence $\gamma^{\rm opt}$ is $N$-independent.

Specifically, for  $W \gg \Omega$, the Leegwater classical-like approximation applies~\cite{leegwater}, and the transfer time is given by~\cite{paperone}
\begin{equation}\label{leeg}
\left\langle\tau_{\text{L}}\right\rangle_W=\frac{N}{\Gamma _{\text{trap}}}+\frac{N\left(N-1\right)}{4 \Omega^2}\left[\gamma +\frac{\Gamma_{\rm trap}}{N}+\frac{W^2}{6\gamma}\left(1-\frac{2 \Gamma_{\rm trap}}{N(2 \gamma+ \Gamma_{\rm trap})} \right)\right] \,.
\end{equation}
The optimal dephasing in this regime is 
\begin{equation}
\gamma^{\rm opt} \approx \frac{W}{\sqrt{6}}\,.
\label{strongW}
\end{equation}

In Fig.~\ref{largeN} we examine explicitly the crossover between the
moderate-disorder regime, where motion is ballistic and $\gamma^{\rm
  opt}$ scales with $N$ in accordance with Eq.~\eqref{midreg}, and the
strong-disorder regime where localization obtains and $\gamma^{\rm
  opt}$ becomes $N$-independent.

\section{Discussion}

An interesting feature of our analysis is the fact that in a
finite chain, dephasing can help transport even in the deep ballistic
regime ($\xi \gg N$), where the spreading of the excitation is very
fast and not only in the localized regime ($\xi \ll N$), as already
discussed in many publications. 

Understanding how dephasing can help transport in the localized regime
is not difficult. Here transport is suppressed in the absence of dephasing, and the excitation spreads only up to a
length $\xi$, whereas nonzero dephasing frees the
excitation leading to a diffusive spreading at large times. 
Thus, in this regime the spreading of the excitation is much
faster in the presence of dephasing than without it. 

\begin{figure}
	\includegraphics[scale=0.7]{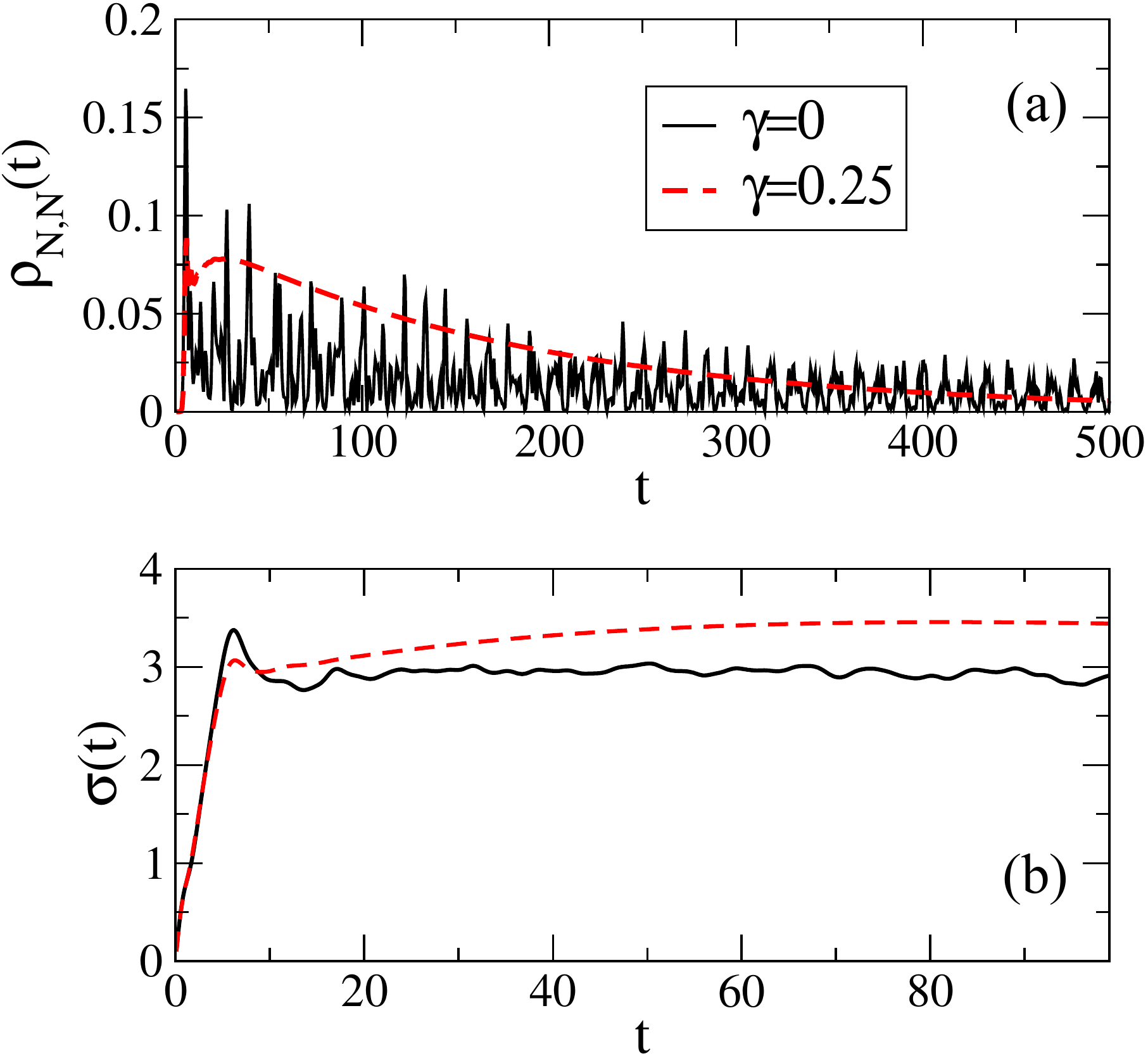}
	\caption{(Color online) (a) Probability to be at site $N$ as a
          function of time, for an excitation initially at site $1$.
           Here $\Omega=1$, $W=2$, $N=10$, and
           $\Gamma_{\text{trap}}=1/16$ as in Fig.~\ref{largeN}. 
      The data refer to a single disorder realization without
      dephasing (solid black curve) and
          at the optimal dephasing, $\gamma \approx 0.25$ (red dashed
          curve). (b) Excitation spreading $\sigma(t)$
          for the same parameters as in panel (a). Here an average is performed over
          $100$ disorder realizations. Note
          that the spreading is defined as:
          $\sigma^2(t)=\sum_n \rho_{n,n}(t) n^2- (\sum_n \rho_{n,n} n)^2 $.
 }\label{rhoNN}
\end{figure}

In the ballistic regime, on the other hand, dephasing can even slow
down the spreading of the excitation
(indeed when dephasing is sufficiently  large, it induces a
diffusive spreading in a long chain which is much slower than the ballistic transport
associated with zero dephasing).
Nevertheless the efficiency of the energy transfer 
depends not only on the rate of excitation spreading, but also
on the probability to be on the last site, which
is coupled to the sink. In the clean case coherences induce large
fluctuations in this probability as shown in
Fig.~\ref{rhoNN}(a), in contrast with the case of nonzero dephasing where the
fluctuations are smoothed so that that on average the excitation
spends more time on the last site, thus increasing the transfer
efficiency. This enhancement due to dephasing can happen even if the
rate of excitation spreading is the same without or with dephasing, as seen in the initial linear growth of the spreading $\sigma(t)$  in
Fig.~\ref{rhoNN}(b).
Of course, too high a rate of dephasing for a given chain length and a given strength of static
disorder will suppress transport,
turning ballistic spreading into diffusive. 
Nevertheless for any
finite size chain, there is always a finite optimal dephasing even in
the ballistic regime. 
 
Finally we would like to stress that in this paper we have focused on the
case of the excitation starting from the first site. Here we
found that dephasing can help only above a critical static disorder. 
On the other hand, starting from other initial conditions, dephasing
can help energy transfer even in the absence of static disorder. 
For instance for a three-site chain with $W=0$ (the clean case), when the
excitation starts from the middle site, the transfer time is
analytically given by
\begin{equation}
   \tau =  \frac{3}{\Gamma_{\rm trap}} +\frac{1}{2\gamma+\Gamma_{\rm
       trap}} + \frac{2\gamma + \Gamma_{\rm trap}}{2\Omega^2} \,,
\label{ave3}
\end{equation}
which gives an optimal dephasing $\gamma^{\rm opt} = (\sqrt{2}\Omega
-\Gamma_{\rm trap})/2$ even in the absence of any static disorder. 
More generally, when the excitation starts in the middle of a clean chain of length $N$, for $N$ odd,
in the limit of small opening $\Gamma_{\rm trap}$  ($\Gamma_{\rm trap} \ll \Omega, \gamma$) we observe that
\begin{equation}
\tau = \frac{N}{\Gamma_{\rm trap}} + \frac{1}{2\gamma}  + \frac{(3 N^2-4 N +1)\gamma}{16\Omega^2} \,, 
\label{middlesitetau}
\end{equation}
and the optimal dephasing is therefore given by
\begin{equation}
\gamma^{\rm opt} = \frac{\sqrt{8}\Omega}{\sqrt{3 N^2-4 N +1}} +O(\Gamma_{\rm trap})\,,
\label{middlesitegam}
\end{equation}
which falls off as $\Omega/N$ for long chains.
A more detailed investigation of the role of noise in the absence of
static disorder and the dependence on initial conditions will be done
elsewhere. For the moment we only comment that
Eqs.~(\ref{ave3}) and (\ref{middlesitetau}) confirm that even in the ballistic regime dephasing
can help energy transfer, in a finite-size system.

\section{Conclusions}
\label{sec-conc}
We have systematically studied the effect of dephasing on transport in
open disordered chains of arbitrary length. Specifically, we have considered
a linear chain of $N$ sites with nearest-neighbor coupling $\Omega$, in the
presence of static disorder of strength $W$ and dephasing of strength
$\gamma$, and where the last site is coherently coupled to an external
environment with opening strength $\Gamma_{\text{trap}}$ and the
system is initialized in the first site.
For this model, which has been extensively studied in the literature, our
analysis allowed us to estimate the critical static disorder above which
dephasing can assist transport.  
Specifically we have seen that $W^{\rm cr}$ varies linearly or inversely with the
opening strength $\Gamma _{\text{trap}}$ when $\Gamma _{\text{trap}}$
is small or large,  respectively.

An essential point of our analysis is the estimate of the 
optimal dephasing rate in
the ballistic regime (and not only in the localized regime as has
been done in previous works). 

Different regimes have been obtained for the behavior of the
optimal dephasing rate $\gamma^{\rm opt}$. For $W$ close to $W^{\rm
  cr}$, we have $\gamma^{\rm opt} \sim (W-W^{\rm cr})/N$, whereas for
$\Gamma_{\text{\rm trap}}/\sqrt{N} \ll W \ll \Omega/\sqrt{N}$, the
optimal dephasing becomes opening-independent and scales as
$\gamma^{\rm opt} \sim W/\sqrt{N}$. In both the weak- and
moderate-disorder regimes, dephasing aids transport even though
eigenstates are delocalized and motion is ballistic. 
This can be explained by the fact that the transfer efficiency is not
only determined by the velocity of excitation spreading but also by
the time the excitation spends in the site coupled to the sink. Since
in the ballistic case the probability to be in the last site
experiences large fluctuations, dephasing can help transport by stabilizing the
excitation on the exit site and increasing its probability to escape.   
Finally, for sufficiently strong disorder, $W \gg \Omega/\sqrt{N}$, the quantum states become localized and the optimal dephasing rate becomes $N$-independent, as has been seen in previous works.

\begin{acknowledgments}

This work was supported
in part by the NSF under Grant No. PHY-1205788 and by the Louisiana Board of Regents under contract LEQSF-EPS(2014)-PFUND-376.
	
\end{acknowledgments}

\appendix
\section{ Analytical Expression for Transfer Time  $\tau_3$ in a 3-site Chain} \label{Appendix}

For a 3-site chain with arbitrary on-site energies $\omega_i$, inter-site hopping $\Omega$, dephasing rate $\gamma$, and site $3$ coupled to the acceptor system with coupling
$\Gamma _{\text{trap}}$, we may solve
Eq.~\eqref{tau_liuov} exactly using Wolfram Mathematica to obtain the transfer time
\begin{equation}
\tau_3=\frac{X_0+\Omega^2 X_2 +\Omega^4 X_4}{Z}\,,
\label{tau3exact}
\end{equation}
where 
\begin{eqnarray}
X_0&=& \Gamma _{\text{trap}} \left[4
	\left(\gamma ^2+(\omega_1-\omega_3)^2\right)+4 \gamma  \Gamma _{\text{trap}}+\Gamma_{\text{trap}}^2\right] \\
	& \times & \left[2 \gamma  \left(3 \gamma^2+\omega_1^2-2 \omega_1\omega_2
		+3 \omega_2^2-4 \omega_2
	\omega_3+2 \omega_3^2\right)+\Gamma_{\text{trap}} \left(5 \gamma ^2+(\omega_1-\omega_2)^2\right)+\gamma  \Gamma_{\text{trap}}^2\right] \,, \nonumber \\
	X_2&=&48 \gamma ^2 \left(\gamma^2+(\omega_1-\omega_3)^2\right)+8 \gamma
	\Gamma _{\text{trap}} \left(15 \gamma ^2+(\omega_1-\omega_3) (5 \omega_1-\omega_2-4\omega_3)\right)  \\
	& + &	 8 \Gamma _{\text{trap}}^2
	\left(11 \gamma^2+(\omega_1-\omega_3)^2\right)
		+24 \gamma  \Gamma _{\text{trap}}^3+2\Gamma_{\text{trap}}^4 \,, \nonumber \\
		X_4 &=& 12 \left(2 \gamma
	+\Gamma _{\text{trap}}\right) \left(4 \gamma +\Gamma
	_{\text{trap}}\right) \,,  \\
	Z&=& 2 \Omega ^2 \Gamma _{\text{trap}}
	\left(2 \gamma +\Gamma _{\text{trap}}\right)
	\left[2 \Omega ^2 \left(4 \gamma +\Gamma_{\text{trap}}\right)+\gamma  \left(2\gamma
	+\Gamma_{\text{trap}}\right)^2+4\gamma(\omega_1-\omega_3)^2\right] \,.
	\end{eqnarray}



\end{document}